\documentclass{optica-article}

\journal{opticajournal} 

\articletype{Research Article}

\usepackage{lineno}

\usepackage{amsmath}
\begin{document}

\title{Narrow linewidth semiconductor lasers based on nonlinear self-injection locking }

\author{Andrew M. Bishop\authormark{1} and  Alexander L. Gaeta,\authormark{1,2,*}}

\address{\authormark{1}Department of Applied Physics and Applied Mathematics, Columbia University, New York, NY 10027, USA\\
\authormark{2}Department of Electrical Engineering, Columbia University, New York, NY 10027, USA}

\email{\authormark{*}a.gaeta@columbia.edu} 


\begin{abstract*}
Self-injection locking techniques for stabilizing lasers have been developed using passive cavities to increase the effective lifetime of the laser cavity, thereby reducing the linewidth of the laser. We propose and demonstrate a new technique based on nonlinear self-injection locking (N-SIL) which we implement via feedback from the gain-narrowed Stokes mode of a fiber Brillouin oscillator. By blue-shifting the Stokes field back to its pump frequency with an electro-optic modulator we realize recursive linewidth reduction that eliminates the phase drift caused by spontaneous emission noise. The fundamental linewidth limit is set by the spontaneous emission limit of the nonlinear oscillator, far lower than the spontaneous emission limit of a semiconductor laser. We demonstrate the power of this approach by achieving sub-hertz fundamental linewidth from the output of a commercial DFB laser and noise performance that significantly exceeds that of conventional SIL. We also and propose alternative fully-integrated designs in CMOS-compatible photonic platforms that allow for highly compact and robust implementations.
\end{abstract*}

\section{Introduction}
Narrow linewidth lasers are essential for high-precision experiments and instrumentation, including atomic clocks, photonic microwave oscillators, gyroscopes, and coherent communications \cite{Jiang2011,AlTaiy,Lai2020,Guan2018,Bai2022,Huynh2012,Pan2013}. The theoretical limit on the laser’s linewidth in the absence of technical noise is given by the Schawlow-Townes linewidth (STL), which is determined by spontaneous emission noise \cite{arthur1958infrared}. In the case of semiconductor lasers, this limit is increased due to the dependence of the refractive index on carrier density as modeled by the Henry factor \cite{henry1982theory}. Creating compact, low-cost lasers for narrow linewidth applications will require highly stable semiconductor lasers, but these lasers on their own typically have natural linewidths that are far too large for precision applications \cite{tran2019tutorial}.

Many methods have been developed for narrowing the linewidth of semiconductor lasers. Extended cavity and distributed Bragg reflector lasers increase the round-trip time of the laser cavity and decrease the cavity resonance bandwidth, which lower the STL of the laser \cite{huang2019high}. Similarly, self-injection locking stabilizes the laser frequency by coupling it to the narrow resonance of a high-Q auxiliary cavity \cite{kondratiev2017self,shim2021tunable,corato2023widely}. Stimulated Brillouin oscillators (SBOs) produce a much lower linewidth signal due to the nonlinear interaction of the pump, signal, and acoustic phonons \cite{stokes1982all,ippen1972stimulated}. This gain narrowing decreases the linewidth of the Stokes signal since the short-lived, low intensity, acoustic wave adjusts its frequency quickly to maintain energy conservation between a noisy pump and a stable Stokes signal \cite{debut2000linewidth,suh2017phonon,gundavarapu2019sub,spirin2021sub}. Active feedback locking techniques such as the Pound-Drever-Hall lock can increase long-term frequency stability by electronically locking the laser to a stable reference cavity, but the frequency noise reduction is limited by the servo locking bandwidth \cite{drever1983laser,black2001introduction,wang2022silicon}. 

In this work, we propose and demonstrate a new technique for producing a narrow linewidth semiconductor laser through a combination of self-injection locking and Brillouin gain narrowing. By taking the narrow signal produced by Brillouin oscillation, shifting its frequency back to the pump frequency, and injecting it into the pump laser, a recursive process begins where noise in the pump laser is eliminated. This technique of injection locking the pump wave of an asymmetric three-wave nonlinear oscillator with its narrowed signal wave can be extended to other systems such as optical parametric oscillators in both $\chi^{(2)}$ and $\chi^{(3)}$ media in fully integrated photonic platforms.

\section{Theory}
Theoretical analysis of semiconductor laser linewidth with external feedback has been significantly discussed for over 40 years. In 1982, Henry \cite{henry1982theory} extended the STL limit to include a linewidth enhancement factor due to changes in carrier density. This work was built upon further to analyze extended cavity diode lasers \cite{agrawal1984line} and resonant feedback \cite{laurent1989frequency}. With the advent of self-injection locking, new formulations were created to describe the linewidth reduction from resonant feedback \cite{kondratiev2017self}. These analyses all use coupled rate-equation models to describe the gain and feedback dynamics which, when linearized, can show the effect of carrier, intensity, and phase noise in the laser on the output frequency.

Similarly, the gain-narrowing effect in SBOs has been extensively modeled and documented. Brillouin scattering is a three-wave interaction converting a pump photon into an acoustic phonon and a lower frequency Stokes photon. By examining the linearized rate equations for this interaction, it can be shown that fluctuations in the pump frequency are transferred to the phonon, isolating the Stokes wave by a factor proportional to the square of the ratio of Stokes wave and phonon lifetimes \cite{debut2000linewidth}. This effect is possible because the short-lived phonon adjusts its frequency to shifts in the pump frequency much faster than the Stokes wave. Brillouin gain narrowing has been used to produce narrow linewidth lasers in both optical fiber systems and integrated microresonators \cite{gundavarapu2019sub,spirin2021sub}. The Brillouin lasing process has a fundamental linewidth limit with a form similar to the STL, where in addition to the thermal quanta in the optical mode (which is typically negligible), thermal quanta are present in the mechanical mode that increase the linewidth limit \cite{suh2017phonon,li2012characterization}.

Self-injection locking can be modeled using the same coupled rate equations that have been used to model both semiconductor lasers with external feedback and Brillouin gain narrowing. Following the notation of Henry \cite{henry1983theory}, the self-injection locking rate equations can be modeled as,
\begin{align}
\dot I_L  &= (G - \gamma) I_L + R + 2\kappa \sqrt{I_L I_M} \cos(\phi_M - \phi_L) + F_I(t),\\
\dot \phi_L&= \frac{\alpha}{2} ( G - \gamma) + \kappa \sqrt{\frac{I_M}{I_L}} \sin(\phi_M - \phi_L) + F_\phi(t),\\
\dot N &= J - S(N) - G I_L + F_N(t),\\
\dot I_M &= -\gamma_M I_M + 2 \kappa_M \sqrt{I_M I_L} \cos(\phi_M - \phi_L),\\
\dot \phi_M &= - \kappa_M \sqrt{\frac{I_L}{I_M}} \sin(\phi_M - \phi_L),
\end{align}
where $I_L, \phi_L$ and $I_M,\phi_M$ are the intensity and phase of the laser and microresonator modes, respectively, $N$ is the population inversion of the gain medium, $J$ is the pump current, $S(N)$ is the carrier recombination rate , $G, \gamma, R$, and $\alpha$ are the laser cavity gain, loss, spontaneous emission rate, and Henry factor, respectively, $\gamma_M$ is the microresonator loss, $\kappa_M$ is the coupling coefficient of the microresonator, and $\kappa$ is the reflection and output coupling from the microresonator back to the laser. Noise in the laser is modeled  with the Langevin noise terms, $F_I, F_\phi$, and $F_N$.

To examine the phase noise in this model, the equations are linearized about their steady-state values and solved using Fourier analysis. The Langevin terms can be grouped as one effective term including fluctuations in phase caused by intensity and population fluctuations. The rate equations for the phase terms of the laser and microresonator modes yield the following simplified equations:
\begin{align}
\dot \phi_L &= - \frac{2 \kappa \kappa_M}{\gamma_M} \sin( \phi_L - \phi_M) + F_\phi(t),\\
\dot \phi_M &= \frac{\gamma_M}{2} \sin(\phi_L - \phi_M),
\intertext{from which a rate equation for the phase difference can be derived to yield,}
\dot \psi &= - \left(\frac{\gamma_M}{2}  + \frac{ 2 \kappa \kappa_M}{\gamma_M}\right) \sin(\psi) + F_\phi(t),
\end{align}
which resembles the Adler equation \cite{adler1946study}. For small perturbations on $F_\phi$, the phase difference will reach a steady-state value, locking the laser and microresonator phases together. This creates a transfer function for the phase-noise Langevin noise term to the laser phase rate equation,
\begin{align}
\dot \phi_L = \left(\frac{\gamma_M^2}{\gamma_M^2 + \kappa \kappa_M}\right) F_\phi(t).
\end{align}

Using the method described in \cite{henry1983theory}, the linewidth can be derived from the phase rate equation above. The coupling and loss rates can be substituted for the cavity quality factors and reflectivity to reach the form,
\begin{align}
\Delta f = \frac{Q_d^2}{Q_M^2 |\Gamma|^2} \frac{R}{4 \pi I_L} =  \frac{Q_d^2}{Q_M^2 |\Gamma|^2}\Delta f_0,
\end{align}
which shows that the linewidth of the self-injection locked laser is reduced from its free-running linewidth by a factor proportional to the square of the ratio of the laser and microresonator quality factors and the strength of the reflection from the microresonator. 

N-SIL combines the feedback-induced stabilization of self-injection locking with the gain narrowing of Brillouin lasing. This process can be modeled using the same rate equation methods as above with the equations representing the laser, pump, signal, and idler mode intensity and phase and the laser population inversion. Since the pump mode is strongly driven by the laser, an adiabatic approximation for the pump mode phase sets it equal to the laser phase. Using this assumption while linearizing and simplifying the equations, the model produces two rate equations for the phase difference between the SBO modes ($\phi_B = \phi_L - \phi_S - \phi_{Ph}$) and the phase difference between the laser and the injected signal ($\phi_{LS} = \phi_L - \phi_S$), such that,

\begin{align}
\dot \phi_B &= -(\gamma_{Ph} + \gamma_{S})\sin(\phi_B) - \kappa'\sin(\phi_{LS}) + F_\phi(t),\\
\dot \phi_{LS} &= - \kappa'\sin(\phi_{LS}) - \gamma_{Ph} \sin(\phi_B) + F_\phi(t),
\end{align}
where $\gamma_S$ and $\gamma_{Ph}$ are the decay rates of the Stokes and phonon modes, respectively, and the coupling constant $\kappa'$ is given by,
\begin{align}
\kappa' = \kappa \left[ I_L \frac{ \gamma_{Ph}}{\gamma_S} \left(\frac{2 \kappa_P}{\beta}\right)^2\right]^{1/4},
\end{align}
where $\kappa$ and $\kappa_P$ are the linear coupling of the laser to the injected field and the pump to the laser field, respectively, and $\beta$ is the strength of the nonlinear coupling between pump, signal, and idler. By diagonalizing the coupling matrix between these two quantities, two separable Adler equations are produced that can be locked simultaneously \cite{adler1946study,pikovsky_rosenblum_kurths_2001}. A new rate equation for the laser phase can be extracted from this system of equations that has the same Adler-like form, which demonstrates that the phase noise produced by the Langevin noise terms will not be transferred to the laser phase because of the feedback from the nonlinear oscillator. \\

\begin{figure}[ht!]
\centering\includegraphics[width=12cm]{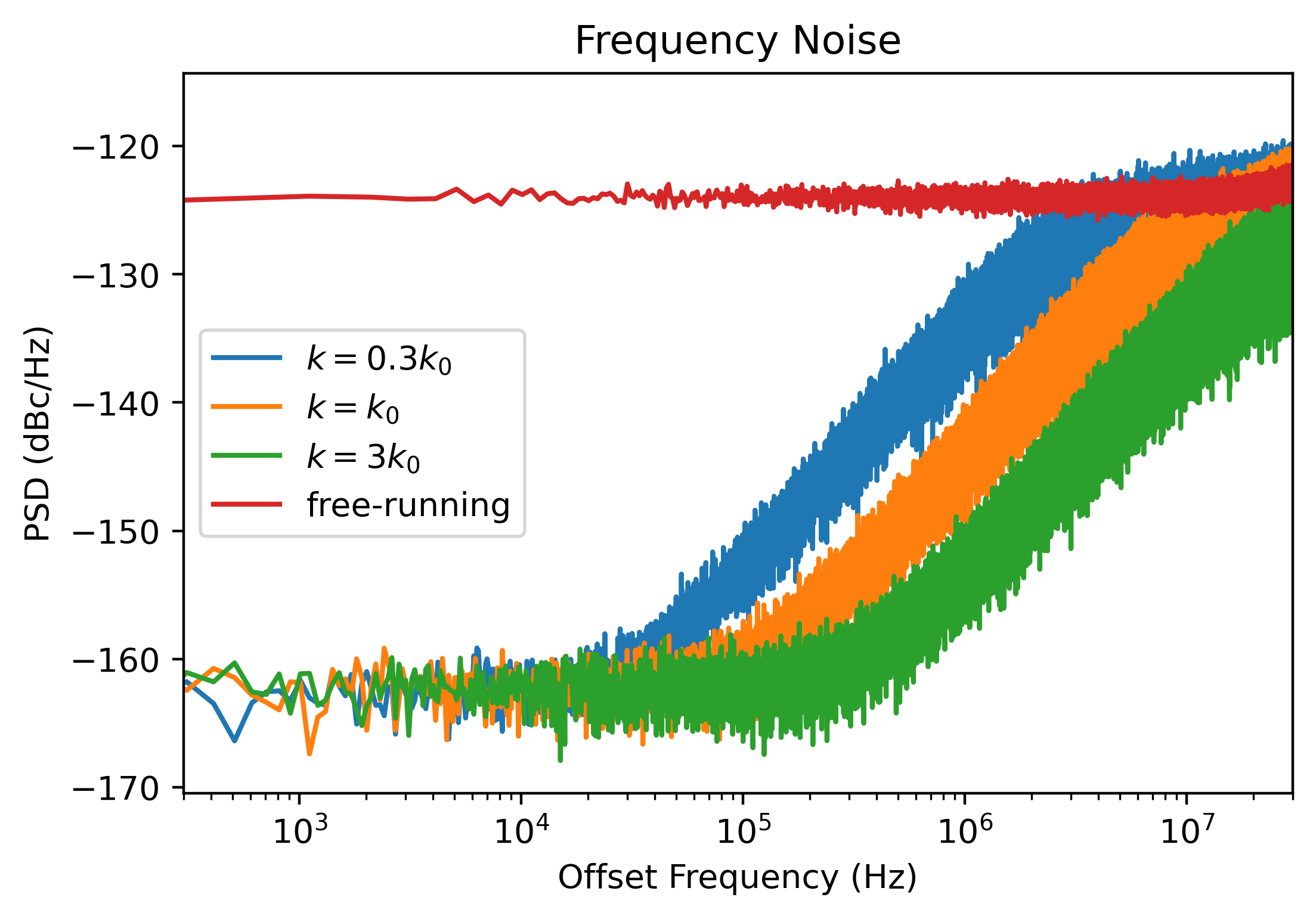}
\caption{Simulated frequency noise power spectral density of the laser output. The feedback strength from the SBO to the laser, $k$, is varied and shows the reduction in frequency noise with increased feedback. }
\label{fig:sim}
\end{figure}

This elimination of the slow phase drift of the laser is observed in simulations of the coupled rate equations model. The laser parameters were drawn from Henry \cite{henry1983theory}, and the equations were propagated using an RK4(5) method with white noise injected into the phase terms to simulate spontaneous emission. After reaching steady-state, the intensity and population inversion were held fixed to reduce the computation time as it was observed including these fluctuations did not have a significant impact on the frequency noise. The simulated frequency noise of the free-running laser and the N-SIL laser are shown in Fig.~\ref{fig:sim}. The locking eliminates the laser's frequency noise within the locking bandwidth, and is instead limited by the spontaneous emission noise from the SBO. The locking bandwidth is determined by the cavity linewidth and feedback strength, similar to the linear self-injection locking case. The frequency noise for varying feedback strength is shown in Fig.~\ref{fig:sim}, demonstrating that increased feedback strength increases the noise reduction bandwidth.

\section{Experiment}

\begin{figure}[ht!]
\centering\includegraphics[width=12cm]{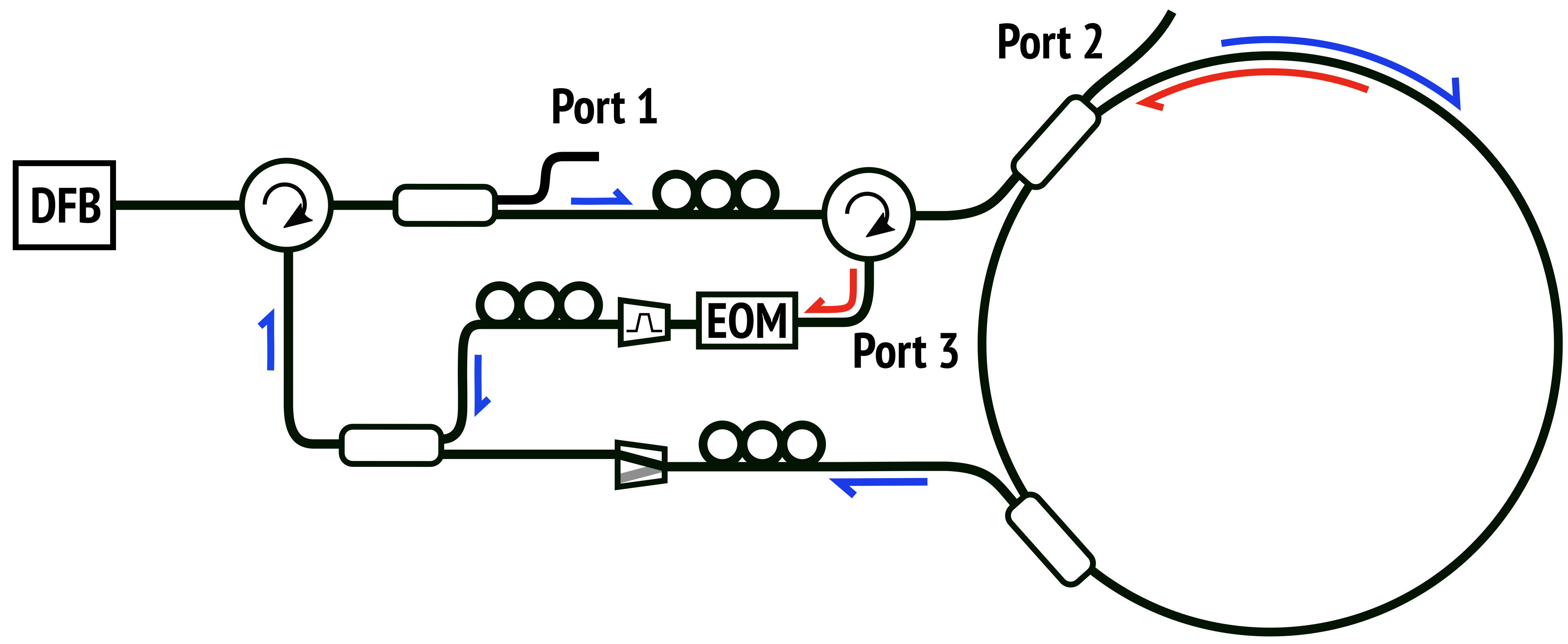}
\caption{Experimental apparatus for nonlinear self-injection locking with a SBO.}
\label{fig:diagram}
\end{figure}

The experimental apparatus for nonlinear self-injection locking is shown in Fig.~\ref{fig:diagram}. The pump laser is a semiconductor DFB laser from QPhotonics driven by a low-noise current source. Its output propagates through two circulators and a polarization controller to reach the Brillouin fiber cavity. The fiber loop consists of 50 m of TrueWave RS fiber, which sits inside an acoustically isolated box. The pump is coupled into the loop through a 90:10 splitter, and a locking signal is coupled out through a 99:1 drop port. This locking signal is fed through the 10\% port of a 90:10 coupler, a polarization controller, and through the first circulator back to the pump laser. 

Inside the fiber loop, the pump generates a counter-propagating SBO signal that is coupled out through the second circulator. This signal is sent to a low insertion-loss amplitude modulator. The modulated signal is filtered to remove the unmodulated SBO carrier and the red-shifted sideband. The blue-shifted sideband is then amplified with an erbium-doped fiber amplifier (EDFA) and injected into the 90\% port of the locking signal coupler to be sent to the pump laser. Nonlinear self-injection locking is initiated by tuning the RF driving frequency of the modulator to the Stokes frequency shift between the pump laser and the Stokes mode of the SBO. The RF driving frequency and the EDFA pump current can be adjusted to change the detuning and feedback strength of the Brillouin locking signal, respectively.

The resulting frequency noise of the output light at Port 1 is measured using a delayed self-heterodyne interferometer (DSHI). The signal being measured is sent to an unbalanced Mach-Zehnder interferometer, where one arm contains an acousto-optic modulator driven by a low-noise RF signal and the other arm contains a 1.6 km fiber delay line in an acoustically isolated box. Upon recombining on the fast photodiode, the frequency-shifted signal from the modulator interferes with the delayed signal to produce a beat note which is measured with a phase-noise analyzer.

\begin{figure}[ht!]
\centering\includegraphics[width=12cm]{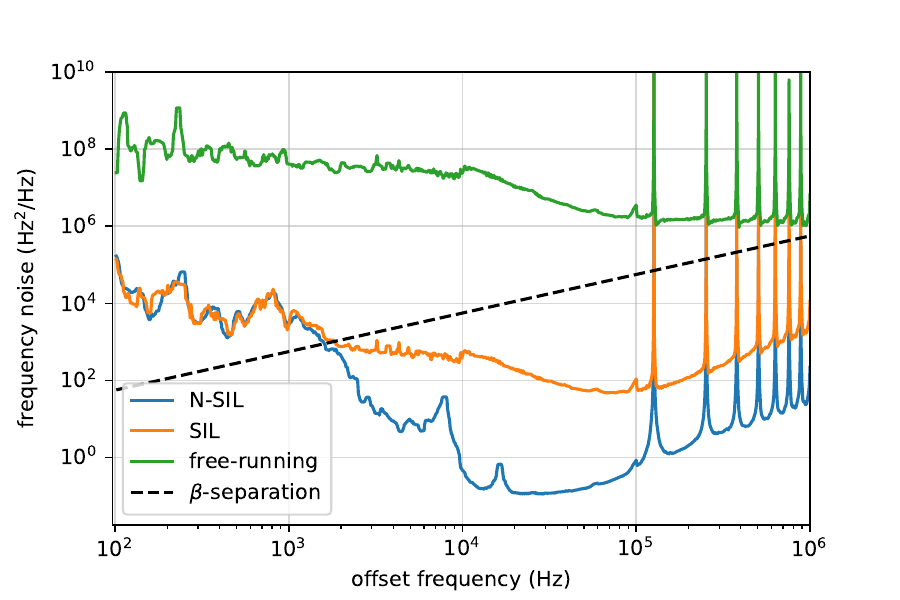}
\caption{Frequency noise power spectral density of the DFB laser measured at port 1 with no feedback (green), self-injection locking (orange), and nonlinear self-injection locking (blue). The dashed line shows the $\beta$-separation line for estimating linewidth.}
\label{fig:PSD}
\end{figure}

The power spectral density of the free-running, self-injection locked, and nonlinear self-injection locked laser signal are shown in Fig.~\ref{fig:PSD}. The self-injection locked signal has a significantly lower frequency noise spectrum compared to the free-running case, but the lock is short-lived because the locking bandwidth is narrow compared to the frequency noise of the laser. Self-injection locking also requires a fixed round-trip phase between the laser and auxiliary cavity to maintain constructive feedback. Active control of this phase offset has been demonstrated in experiments with self-injection locking to fiber loops but was not implemented in this experiment \cite{spirin2021sub}.

The nonlinear self-injection locking power spectral density shows greater noise reduction than the self-injection locked case, with an integrated linewidth of 8 kHz, and a fundamental linewidth of 0.35 Hz. As the injection locking narrows the pump laser, this noise reduction is recursively passed back to the Brillouin oscillator. The noise reduction at low offset frequencies is limited by the phase stability of the fiber cavity, whose thermal and mechanical noise are imprinted on the Brillouin Stokes mode. Unlike in the linear self-injection locking case, phase drift accumulated along the injected signal path does not affect the lock. While this noise will be passed to the pump laser, it will be removed by the stimulated Brillouin scattering and therefore not disrupt the lock.

\begin{figure}[h!]
\centering\includegraphics[width=10cm]{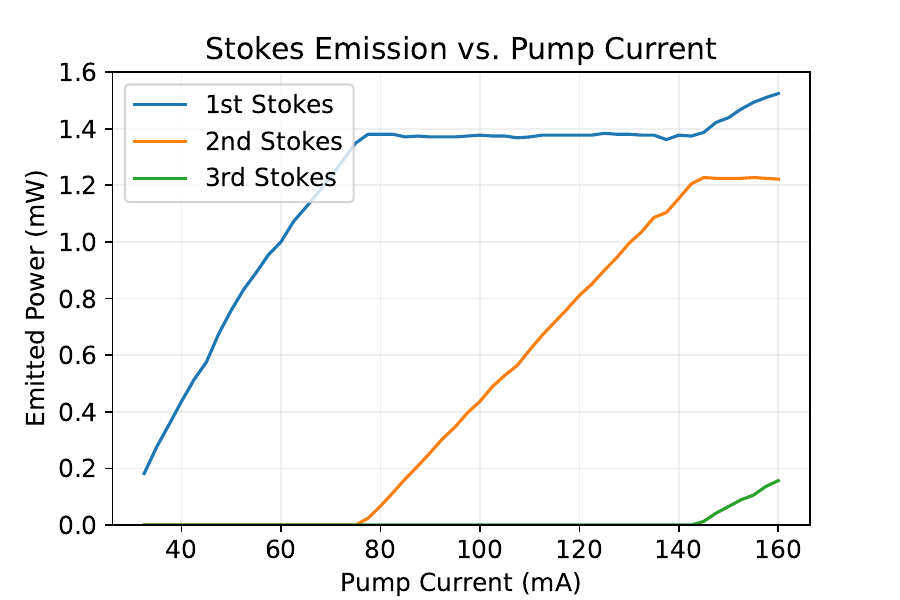}
\caption{Power in each Stokes mode measured with varying pump current.}
\label{fig:cascade}
\end{figure}

The operating conditions of the Brillouin oscillator and the feedback path determine the stability and noise reduction of the nonlinear self-injection lock. The Brillouin oscillator must remain above threshold with enough power in the Stokes mode to inject into the pump laser. The Brillouin oscillator can also operate above the threshold for cascaded Brillouin oscillation, such that the first-order Stokes mode power is clamped at threshold and its fundamental linewidth increases with pump power \cite{behunin2018fundamental}. The cascaded thresholds are shown in Fig.~\ref{fig:cascade}, which plots the power in the first three orders of Stokes modes with pump current. From these data, the Brillouin oscillation threshold is found to be 23.3 mA. The frequency noise of the Brillouin signal while the laser was under nonlinear self-injection locking was measured to determine the phonon-limited linewidth of the Brillouin oscillator. At most pump powers, the white-noise floor indicative of a phonon-limited linewidth was below the thermorefractive noise and detector noise floor, making it impossible to measure except at pump powers immediately above threshold. The phonon-limited linewidth at a pump current of 28 mA was measured to be 0.32 Hz, which is consistent with theoretical predictions \cite{suh2017phonon}. At the optimal point just below the first cascaded order threshold of 76 mA, the phonon-limited linewidth is predicted to be 10.9 times lower than at 28 mA, which yields a fundamental linewidth of 29 mHz.

The linewidth of injection-locked lasers have been studied widely both theoretically and experimentally \cite{henry1985locking,spano1986frequency,gallion1985contribution,hui1991injection,ratkoceri2018injection}.  In these analyses, it is found that the locked laser will inherit the phase noise properties of the injected signal. The locking bandwidth increases with injected power up to a point where it induces bistability. The reduction in the locked laser linewidth is continuous  with injected power; as the power increases, the locked laser linewidth moves smoothly from its free-running linewidth to the injected signal linewidth \cite{mogensen1985fm}. Once the locking bandwidth exceeds the phase noise bandwidth of the locked laser, the linewidth reduction saturates at the injected signal linewidth. 

\begin{figure}[ht!]
\centering\includegraphics[width=10cm]{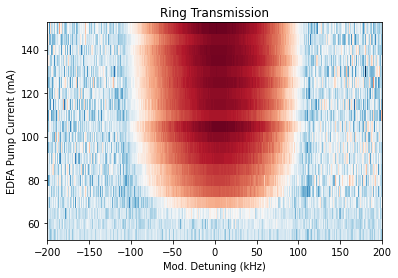}
\caption{Transmitted power at Port 2 measured with varying Brillouin feedback strength tuned with the EDFA current and frequency detuning from the EOM driving signal.}
\label{fig:arnold}
\end{figure}

In the nonlinear self-injection locked system, the locking bandwidth can be measured by tuning the RF driving frequency of the EOM. This shifts the pump-Stokes frequency offset, effectively pulling the pump laser across the fiber cavity resonance. The pump-cavity detuning can be observed by measuring the transmission past the input coupler of the cavity at Port 2 in Fig~\ref{fig:diagram}. Tracking the locking bandwidth with injected signal power produces an Arnold tongue, shown in Fig.~\ref{fig:arnold}\cite{pikovsky_rosenblum_kurths_2001}. When the injected power increases, the lock is initiated and then the locking bandwidth increases. This effect saturates when the locking bandwidth reaches the resonance bandwidth of the fiber cavity, past which the laser cannot be pulled without stimulated Brillouin oscillation ceasing. The frequency noise at high offset frequencies above 100 kHz decreases as the injected power increases, and that plateaus once the locking bandwidth reaches the cavity resonance bandwidth. It is possible if the cavity resonance limit on the locking bandwidth were lifted, the frequency noise could decrease further, as the Brillouin injected signal at this offset frequency is much lower than the laser output as measured at Port 1.

\section{Conclusions}
We have analyzed and implemented a new mechanism for reducing the linewidth of semiconductor lasers through self-injection locking to a nonlinear oscillator. In our demonstration, a noisy semiconductor DFB laser drives a fiber Brillouin oscillator such that the Stokes signal is used to lock the laser to the fiber cavity resonance and significantly reduce its linewidth. The linewidth of the nonlinear self-injection locked laser is shown to be narrower than the self-injection locked configuration of the same pump laser and fiber cavity. The process of nonlinear self-injection has also been explored theoretically to show its improvements over linear self-injection locking in removing spontaneous emission noise inherent in semiconductor lasers. This theory is also not limited to the Brillouin fiber oscillator platform and can be applied to a variety of optical systems where asymmetric nonlinear oscillators are found. These include a variety of well-developed photonic platforms which could be used to implement an integrated design for nonlinear self-injection locking. Stimulated Brillouin oscillators have been demonstrated on silicon and silicon nitride photonic platforms and could support a similar mechanism as the demonstration here \cite{gundavarapu2019sub}\cite{otterstrom2018silicon}.  Optical parametric oscillators (OPOs) with asymmetric losses between their signal and idler modes have been theoretically examined for applications to the reduction of phase noise \cite{gentry2016passive}. A nonlinear self-injection locking device could be implemented on either a $\chi^{(2)}$ platform with second harmonic generation and a degenerate OPO or a $\chi^{(3)}$ platform with cascaded OPOs to generate the gain-narrowed signal and reconvert it to the pump laser’s frequency. This theory of nonlinear self-injection can be useful for improving the frequency stability of low-precision lasers for many applications and across many platforms.

\begin{backmatter}
\bmsection{Funding}
 This work was Army Research Office (ARO) (Grant No. W911NF-21-1-0286) and Air Force Office of Scientific Research (AFOSR) (Grant No. FA9550-20-1-0297).
\end{backmatter}

\bibliography{NSIL}

\end{document}